\documentclass[twoside,fleqn]{article}
\usepackage{espcrc2}
\usepackage{graphicx}

\newcommand{\be}{\begin{equation}}
\newcommand{\ee}{\end{equation}}
\newcommand{\bea}{\begin{eqnarray}}
\newcommand{\eea}{\end{eqnarray}}

\newcommand{\nn}{\nonumber}

\newcommand{\cO}{{\cal O}}
\newcommand{\AmS}{{\protect\the\textfont2
  A\kern-.1667em\lower.5ex\hbox{M}\kern-.125emS}}

\title{Pad\'e Unitarizations: A Critical Look}

\author{Javier Virto \address{ Universit\`a di Roma ``La Sapienza'' and INFN Sezione di Roma, 00185 Roma, Italy.} \thanks{I would like to thank the organizers of QCD'08 for a very interesting workshop, as well as P.Masjuan and J.J.Sanz-Cillero for a pleasant collaboration. I would also like to acknowledge positive criticism from J.R.Pel\'aez and G.Rios.} }

\begin{document}

\begin{abstract}
We use the Linear Sigma Model to test the convergence of different sequences of Pad\'e Approximants when used to unitarize the low energy $\pi\pi$ scattering amplitude. We find that, in this particular case, diagonal sequences reproduce the sigma pole with high accuracy, as opposed to other sequences that have been discussed extensively.
\end{abstract}

\maketitle

\section{INTRODUCTION}

One of the most interesting aspects of strong interactions is the enormously rich and complicated hadronic spectrum that arises from such a simple high energy structure. How does exactly this spectrum arise from QCD is a question under study. The understanding of QCD at low energies is basically provided by ChPT \cite{ChPT}, the low energy effective theory of QCD, which gives also some insight on the spectrum issue. Although ChPT breaks down in the vicinity of the resonance region, arguments following from unitarity and analyticity are able to extract information on some resonances from the chiral Lagrangian, which contains such information hidden in the values of its low energy constants. One of the most successful of such \emph{unitarization} procedures is the Inverse Amplitude Method (IAM) \cite{IAM}, which is able to extract with acceptable accuracy the lowest lying scalar and vector resonances.

These proceedings are based on the work of Ref.~\cite{MSV}, where some of the unitarization techniques are tested by means of the Linear Sigma Model, for which exact information about the sigma resonance is available. We focus on the IAM and also on some sequences of Pad\'e Approximants (PAs) \cite{PAs}.

\section{UNITARIZATION OF THE NLSM}

\label{S2}

In order to determine the scalar meson mass and width up to ${\cO}(g)$, we compute the one-loop sigma correlator,
$$
\Delta(s)^{-1}=s\, -\, M_\sigma^{2} \, \bigg[1 + \frac{3 g}{16\pi^2}\, \bigg( -\frac{13}{3} +  \ln\frac{-s}{M_\sigma^2}\quad
$$
\begin{equation}
\qquad+ 3 \rho(s) \ln{\left( \frac{\rho(s)+1}{\rho(s)-1}\right)}  \bigg) + {\cO} (g^2)\bigg]\, ,
\end{equation}
where $\rho(s)\equiv \sqrt{1- 4 M_\sigma^2/s}\ $,  and the term  $-13/3$ is determined by the renormalization scheme that
sets the relation $2 g F^2=M_\sigma^2$ at the one-loop order, with $F$ the pion decay constant. Now it is possible to extract the pole $s_p$ of the propagator up to the considered order in perturbation theory. If one approaches the branch cut from the upper part of the complex $s$--plane, the pole in the second Riemann sheet is located at
\begin{equation}
s_p =
\, M_\sigma^{  2} \,\left[\, 1\, +   \frac{3 g}{16\pi^2}\,
\left( -\frac{13}{3} +  \pi\sqrt{3}  \,  - \,  i \pi  \right) \right]
\end{equation}
up to corrections of $\cO(g^2)$. The pole mass and width, defined from $s_p=(M_p-i \Gamma_p/2)^2$, are then given by
$$
\frac{M_p^2}{M_\sigma^{  2} } = 1 \, +   \frac{3 g}{16\pi^2}\, \left( -\frac{13}{3} +  \pi\sqrt{3}  \right)+ \cO(g^2)\,\qquad
$$
\begin{equation}
\frac{M_p \Gamma_p}{M_\sigma^{  2}} = \frac{3 g }{16\pi} \,+\, \cO(g^2) \, .
\label{eq.LSM-pole}
\end{equation}
We now consider the LSM at low energies. The contribution from the sigma exchanges to the renormalized $\cO(p^4)$ $\chi$PT couplings gives~\cite{ChPT}
\begin{eqnarray}
\ell_1^r(\mu) &=& \frac{1}{4 g } + \frac{1}{96\pi^2}\left[ \ln\frac{M_\sigma^2}{\mu^2} -\frac{35}{6} \right]  \, +\, \cO(g) \, , \nonumber \\[2mm]
\ell_2^r(\mu) &=& \frac{1}{48\pi^2}\left[ \ln\frac{M_\sigma^2}{\mu^2}
-\frac{11}{6} \right]  \, +\, \cO(g) \, .
\label{eq.LSM-LECs}
\end{eqnarray}
The $\pi\pi$--scattering is determined by the $\pi^+\pi^- \to \pi^0\pi^0$ amplitude, which is given up to $\cO(p^4)$ in the chiral expansion  by
$$
A(s,t,u) = \frac{s}{F^2} \, +\, \frac{2 s^2}{F^4} \ell_1^r \, +\, \frac{s^2 + (t-u)^2}{2 F^4} \ell_2^r
$$
$$
+\frac{1}{96\pi^2 F^4} \left[ -3 s^2 \ln\frac{-s}{\mu^2} - t(t-u) \ln\frac{-t}{\mu^2}\right.
$$
$$
\left.-u(u-t) \ln\frac{-u}{\mu^2} + \frac{5 s^2}{2} +\frac{ 7 (t-u)^2}{6} \right]\, ,
$$
where $\mu$ refers here to the arbitrary renormalization scale, and the chiral limit has been considered. With this one constructs the definite isospin partial waves, $t^I_J(s)$, with $IJ=00,11,20$, finding the following $\cO(p^2)$ amplitudes,
$$
t_0^0(s)_{(2)} = \frac{s}{16\pi F^2} \, , \qquad
t_1^1(s)_{(2)} = \frac{s}{96\pi F^2}\, , \qquad\qquad
$$
\begin{equation}
t_0^2(s)_{(2)} = -\frac{s}{32\pi F^2}\, ,
\label{eq.t2}
\end{equation}
and at $\cO(p^4)$,
\begin{equation}
t_0^0(s)_{(4)}\, = \, t_0^0(s)_{(2)} \, \, \times\,\, \frac{11 s}{6 M_\sigma^{  2}}
\label{eq.t4}
\end{equation}
{\flushright{$\displaystyle\times \left[  1  - \frac{g}{264\pi^2} \left(18\ln\frac{-s}{M_\sigma^{  2}} + 7 \ln\frac{s}{M_\sigma^{  2}} +\frac{193}{3}\right)  \right] \, , $}\\}
{\flushleft{$\displaystyle t_1^1(s)_{(4)} \, =\, t_1^1(s)_{(2)} \,\, \times\,\, \left(\frac{-s}{M_\sigma^{  2}}\right)\,\, $}\\}
{\flushright{$\displaystyle\times\left[1  + \frac{g}{48\pi^2} \left(\ln\frac{-s}{M_\sigma^{  2}} - \ln\frac{s}{M_\sigma^{ 2}} -\frac{26}{3}\right)  \right] \, ,$}\\}
{\flushleft{$\displaystyle t_0^2(s)_{(4)} = t_0^2(s)_{(2)} \,\,\times\,\, \left( \frac{- 2 s}{3 M_\sigma^{  2} } \right)\,\, $}}\\[2mm]
$$\displaystyle\quad\times\left[  1  - \frac{g}{24\pi^2} \left(\frac{9}{4} \ln\frac{-s}{M_\sigma^{  2}} + \frac{11}{4} \ln\frac{s}{M_\sigma^{  2}} +\frac{163}{24}\right) \right] \, ,$$\\

\noindent up to corrections of $\cO(g^2)$. Given the $\cO(p^2)$ and $\cO(p^4)$ $\chi$PT amplitudes from Eqs.~(\ref{eq.t2})--(\ref{eq.t4}), it is then possible to extract the poles of the corresponding $t(s)_{_{\rm IAM}}$ for the LSM, satisfying $t(s)_{(4)}/t(s)_{(2)}=1$ at $s=s_p^{IJ}$:\\[3mm]
$$
s_p^{00}= \frac{ 6}{11}  M_\sigma^{  2} \left[  1 \!+ \! \frac{g}{264\pi^2}\left( \frac{193}{3} \!+\! 25 \ln\frac{6}{11}\!\! -\!\!18 i \pi \right)\right] ,
$$
{\flushleft $\displaystyle
s_p^{11}= - M_\sigma^{  2} \left[1  +  \frac{g}{48\pi^2} \left(   \frac{26}{3} + i\pi \right)\right] ,
$}\\[2mm]
$$
s_p^{20} = -\frac{3}{2} M_\sigma^{  2} \left[  1  + \frac{g}{24\pi^2} \left( \frac{163}{24}+5 \ln\frac{3}{2}+ \frac{11 i \pi}{4}\right)\right] .
$$
These are the poles that appear in the unphysical Riemann sheet as one approaches from upper half of the first Riemann sheet. There is also a
conjugate pole at $s_p^*$ if one approaches the real $s$--axis from below.

The first thing to be noticed is that poles appear in the $IJ=11$ and $20$ channels even for small values of $g $, contrary to what one expects in the LSM, where no meson  with these quantum numbers exists. Furthermore, these ``states'' are not resonances, as they are located on the left-hand side of the complex $s$--plane, out of the physical Riemann sheet, and carrying a negative squared mass.

As for the  $IJ=00$ channel, one finds a resonance with pole mass and width,
\begin{eqnarray}
\frac{M_p^2}{M_\sigma^2}&=&  \frac{6}{11} \, \, \left[  1  +  \frac{g}{16\pi^2} \left( \frac{50}{33} \ln\frac{6}{11} +\frac{386}{99}\right) \right] \, \, , \nn \\[2mm]
\frac{M_p \Gamma_p}{M_\sigma^2}  &=& \frac{24}{121}\,\cdot \,  \frac{ 3 g}{16\pi}\,\, +\, \cO(g^2)  \, .
\end{eqnarray}
The IAM predictions for $M_p^2$ and $M_p \Gamma_p$ result, respectively, 40\% and 80\% smaller than the original ones in the LSM, computed in
Eq.~(\ref{eq.LSM-pole}). Therefore, this provides an example of a situation in which this particular method cannot reproduce the hadronic properties
of the theory from its effective low-energy description.

\section{HIGHER ORDER PA's}

Now we consider higher order PAs. In order to be able to handle the amplitude at higher orders, we will consider the $\pi\pi$ scattering at tree-level. This is equivalent to working in the limit $g \ll 1$ and keeping just the first non-trivial contribution in the $g$ expansion.  At low energies the amplitude becomes
\begin{equation}
A(s,t,u)\,\, =\,\, \frac{s}{F^2}\,\left[ 1\, +\, \frac{ s}{M_\sigma^2} \, +\, \frac{ s^2}{M_\sigma^4} \, \, +\,\, ...\right]\, ,
\end{equation}
and the partial waves are given by,
\begin{eqnarray}
t_0^0(s)  &\!\!\!\!=\!\!\!\!& \frac{s}{16 \pi F^2} \, \left[ 1 + \frac{11 s}{6 M_\sigma^2} +\frac{15 s^2}{12 M_\sigma^4}\,\,+\,\,...\right] \, , \nn\\[2mm]
t_1^1(s) &\!\!\!\!=\!\!\!\!&\frac{s}{96\pi F^2} \, \left[ 1- \frac{s}{M_\sigma^2} +\frac{ 9 s^2}{10 M_\sigma^4} \,\,+\,\, ... \right] \, , \nn\\[2mm]
t_0^2(s) &\!\!\!\!=\!\!\!\!&- \frac{s}{32 \pi F^2}\, \left[ 1 - \frac{ 2 s}{ 3 M_\sigma^2} + \frac{s^2}{ 2 M_\sigma^4}\,\, +\,\,...\right] .
\end{eqnarray}

In this section we compare the $[1/N]$ and $[N/N]$ sequences for the study of the $\pi\pi$ partial  wave scattering amplitudes. We shall focus  on the $IJ=00$ partial wave, but analogous results are found for the other channels. Former works pointed out that the PAs and other unitarizations fail to incorporate the crossed channel resonance exchanges. Nonetheless, we will see that as $N$ grows, the poles of the sequence $[N/N]$ actually tend to mimic not only the $s$--channel poles but also the left-hand cut contribution from diagrams with resonances in the $t$ and $u$ channels.

\begin{figure}
\begin{center}
 \includegraphics[width=7cm,height=4cm]{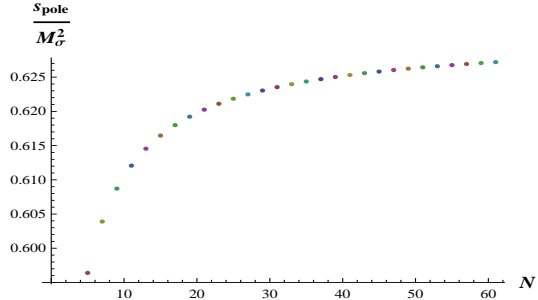}
  \vspace{-0.5cm}
  \caption{Position of the nearest pole to $M_\sigma^2$
  for the first PAs of the form $[1/N]$ with $N$ odd
  (for even $N$ all the poles are complex).}
    \vspace{-1cm}
  \label{LSM30}
\end{center}
\end{figure}

\begin{figure}
\begin{center}
 \includegraphics[width=7cm,height=4cm]{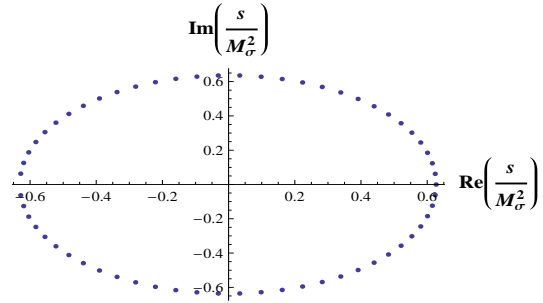}
\vspace{-0.5cm}
  \caption{Poles of the $P^1_{61}$ in the complex plane.}
  \vspace{-1cm}
  \label{LSM302}
\end{center}
\end{figure}

We begin with the sequences of the type $[1,N]$.  Our results are summarized in Figs.~\ref{LSM30} and \ref{LSM302}: No convergence is found with this sequence. In the case of $N$ odd, Fig.~\ref{LSM30} shows that the $P^1_N$ pole closest to $M_\sigma^2$ does not approach this value even for very large $N$, always remaining a 30\% below. The analytical structure of the original amplitude ($s$--channel sigma pole plus left-hand cut) is never recovered since the $[1/N]$ PAs always set the poles in the circular pattern  shown in Fig.~\ref{LSM302}. This  suggests that the use of further $[1/N]$ approximants to extend the IAM is not the optimal way to proceed, even if  we had an accurate knowledge of the low-energy  expansion up to very high orders.

\begin{figure}
\begin{center}
  \includegraphics[width=7cm,height=4cm]{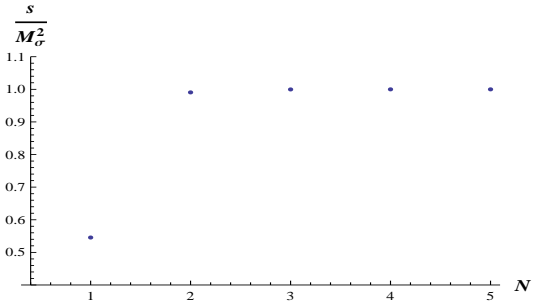}
  \vspace{-0.5cm}
  \caption{
  Location of the closest pole to $M_\sigma^2$
  for the first [N/N] PAs.
  }
    \vspace{-1cm}
\label{PadeNN20}
\end{center}
\end{figure}

Alternatively,  the use of sequences such as $[N+K/N]$ (e.g. $[N-2/N],\, [N-1/N], \, [N/N]\ldots$) seems to be a better strategy. In the following we analyze the sequence $[N/N]$, as it ensures the appropriate behavior  at high energies, $|t(s)|<1$. Nevertheless, similar results have been generally found for the $[N+K/N]$ PAs with $K\neq 0$. The $P^N_N$ pole closest to $M_\sigma^2$ is shown in Fig.~\ref{PadeNN20}. One finds a quick convergence of the sequence: $P_1^1$ reproduces the sigma pole a $40\%$ off but $P^2_2$ disagrees by less than $1\%$, $P^3_3$ by less than $0.1\%$, etc. Notice that already  $P_2^2$ provides a much better description than $P^1_{61}$, although one includes far more low-energy information in the latter. All this points out the sizable discrepancy of the first element of the sequence ($P^1_1$) with respect to the original amplitude. It also indicates that the $[1/N]$ PAs do
not produce a serious improvement. On the contrary, the $[N/N]$ sequences provide a more efficient strategy with a quick convergence.

\begin{figure}
\begin{center}
  \includegraphics[width=7cm,height=4cm]{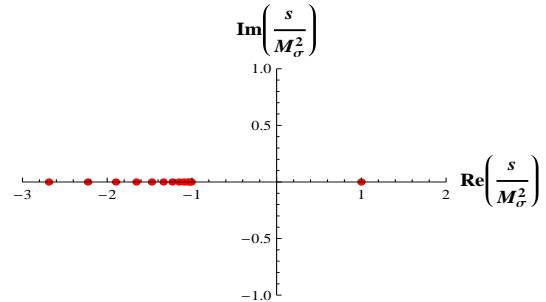}
 \vspace{-0.5cm}
  \caption{
  Poles of $P^{20}_{20}$.
  }
   \vspace{-1cm}
\label{PadeNN202}
\end{center}
\end{figure}

Likewise,  Fig.~\ref{LSM302} shows how the $[1/N]$ PAs  are unable to recover the analytical structure of the original amplitude, whereas the $[N/N]$ sequence, besides providing the isolated pole of the sigma, tends to reproduce the left-hand cut as $N$ increases. The poles of $P_{20}^{20}$ are plotted in Fig.~\ref{PadeNN202}. Although a PA is a rational function without cuts, these are mimicked by placing poles where the cuts should lie. The $P_{20}^{20}$ has one isolated pole near $M_{\sigma}^2$ (with an accuracy of $10^{-30}$) and nineteen poles over the real axis at $s_p<-M_{\sigma}^2$, i.e. on the left-hand cut of the original function. As $N$ is increased, the number of poles lying on the branch cut increases too.

As an amusement, we have also probed the PA sequence $[N/1]$ which has the same number of inputs as the $[1/N]$ has for a given $N$. In this new case we have found convergence in both LSM and the resonance model presented in the following section but slower than the $[N/N]$. For instance, the prediction for the $M_{\sigma}^2$ for the first $P_1^N$ are $\frac{s_p}{M_{\sigma}^2}=0.55,1.47,0.73,1.27,0.81...$ A criticism that can be done to this sequence is its lack of unitarity, in contrast to the other studied sequences.

\section{CONCLUSIONS}

The conclusion seems to be two-fold. First, the LSM features a low energy behavior for which the IAM does not provide a good approximation, at least in what concerns the sigma pole. Second, while some PAs converge amazingly well, even beyond expectation (there are in these cases no theorems ensuring convergence), as for example the diagonal sequences $[N,N]$, some other sequences such as $[1,N]$ show a particularly poor convergence. In addition, the convergent sequences display an analytic structure that mimics perfectly poles and cuts.

Concerning the first point, some explanation has been provided by some of the authors of \cite{IAM}. They argue that the IAM is a good approximation in strongly interacting regimes where unitarity  violations are relevant, while in the case of a weakly interacting set-up as the one described here, in the limit $g\ll 1$ the chiral expansion breaks down. This is due to the fact that in this limit the contact  $\cO(p^4)$ terms are as important as the loop contributions. The IAM would behave properly for values of the LECS of order $10^{-3}$, as is the case in QCD. Moreover, they argue that in the weakly interacting limit the sigma resonance is almost decoupled, and thus the fact that the IAM can pin it down with a precision of about $40\%$ is remarkable. A more careful investigation of how the approximations involved in the derivation of the IAM, break down in the LSM, would help to clarify this issue.

\end{document}